\documentclass[showpacs,twocolumn,floatfix,aps,prl,a4paper]{revtex4}

\usepackage{amsmath}
\usepackage{graphicx}
\usepackage{epsfig}
\usepackage{latexsym,epsfig,bm,times,psfrag,subfigure}
\usepackage{bm,times}
\usepackage{dcolumn}

\begin{document}

\title{Exactly Solved Model for an Electronic Mach-Zehnder Interferometer}
\author{D.\ L.\ Kovrizhin$^{1,2}$ and J.\ T.\ Chalker$^{1}$}
\affiliation{$^{1}$Theoretical Physics, Oxford University, 1, Keble Road, Oxford, OX1
3NP, United Kingdom}
\affiliation{$^{2}$RRC Kurchatov Institute, 1 Kurchatov Sq., Moscow, 123182, Russia}
\date{\today}

\begin{abstract}
We study nonequilibrium properties of an electronic Mach-Zehnder
interferometer built from integer quantum Hall edge states at filling
fraction $\nu{=}1$. For a model in which electrons interact only when they
are inside the interferometer, we calculate exactly the visibility and phase
of Aharonov-Bohm fringes at finite source-drain bias. When interactions are
strong, we show that a lobe structure develops in visibility as a function
of bias, while the phase of fringes is independent of bias, except
near zeros of visibility. Both features match the results of
recent experiments [Neder \textit{et al.} Phys. Rev. Lett. \textbf{96},
016804 (2006)].
\end{abstract}

\maketitle

Questions about phase coherence in interacting quantum systems out of
equilibrium are of fundamental and wide-ranging importance. Despite great
progress over the past decade, many aspects of
nonequilibrium problems remain poorly understood. One recent example of this situation is the
\textquotedblleft unexpected behaviour\textquotedblright\ observed in
state-of-the-art experiments on electronic Mach-Zehnder interferometers
(MZIs) \cite{heiblum2, roulleau07,bieri08} driven out of equilibrium by an
applied bias voltage. In these experiments the visibility of Aharonov-Bohm (AB) fringes in the
conductance shows a lobe--like structure as a function of bias, while the
phase of oscillations is independent of bias even with different interferometer arm lengths, except at zeros of the
visibility where it jumps by $\pi $. 

These observations have attracted a lot of attention. It was
immediately appreciated \cite{heiblum2} that they lie outside
a single-particle description. Moreover, since integer quantum Hall
edge states scale to non-interacting chiral Fermi gases at low energy,
the {\em finite-range} of electron-electron
interactions seems to be crucial. The effort to understand interaction effects in MZIs
at integer filling is therefore linked with work on non-linear effects
in non-chiral Luttinger liquids \cite{glazman}, as well as to interferometry
of fractional quantum Hall quasiparticles \cite{fractional}.
The most obvious consequence anticipated from interactions is dephasing. This may arise from
external noise \cite{marquardt04} or internally \cite{buttiker01,chalker07},
but in both cases is expected to suppress AB fringe visibility smoothly with
increasing bias, in contrast to observations.  It has been found, however, that zeros in
visibility can arise if the edge channels that form the interferometer arms
are coupled to another channel: such an extra channel may be a feature of
sample design \cite{sukhorukov07}, and is present intrinsically at $\nu{=}2$ 
\cite{sukhorukov08}. Although those results are encouraging, they do not seem
sufficiently universal to explain all current experiments. In this context,
two recent papers \cite{neder08,sim08} that obtain visibility oscillations
from calculations of interaction effects at $\nu{=}1$ represent an
interesting advance. These papers contain illuminating physical
insights, and similar phenomena have been shown to exist in another context
\cite{marquardt08},  but approximations used in \cite{neder08,sim08} are not standard ones and their
reliability is hard to judge.

\begin{figure}[tbp]
\epsfig{file=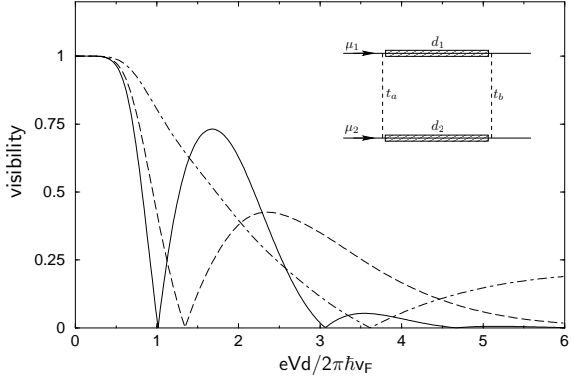,width=7.5cm,angle=0}
\caption{Visibility as a function of bias voltage for MZI with
  $d_1=d_2$ and $t_a^2=t_b^2=1/2$ at
  interaction strengths: $\tilde{\protect\gamma}=1$ (dot-dashed line),
  $\tilde{\protect\gamma}=2$ (dashed line), $\tilde{\protect\gamma}=3$
  (full line), where  $\tilde{\protect\gamma}=2\protect\pi
  \protect\gamma $. Inset: schematic view of model studied.}
\label{fig2}
\end{figure}

\begin{figure}[tbp]
\epsfig{file=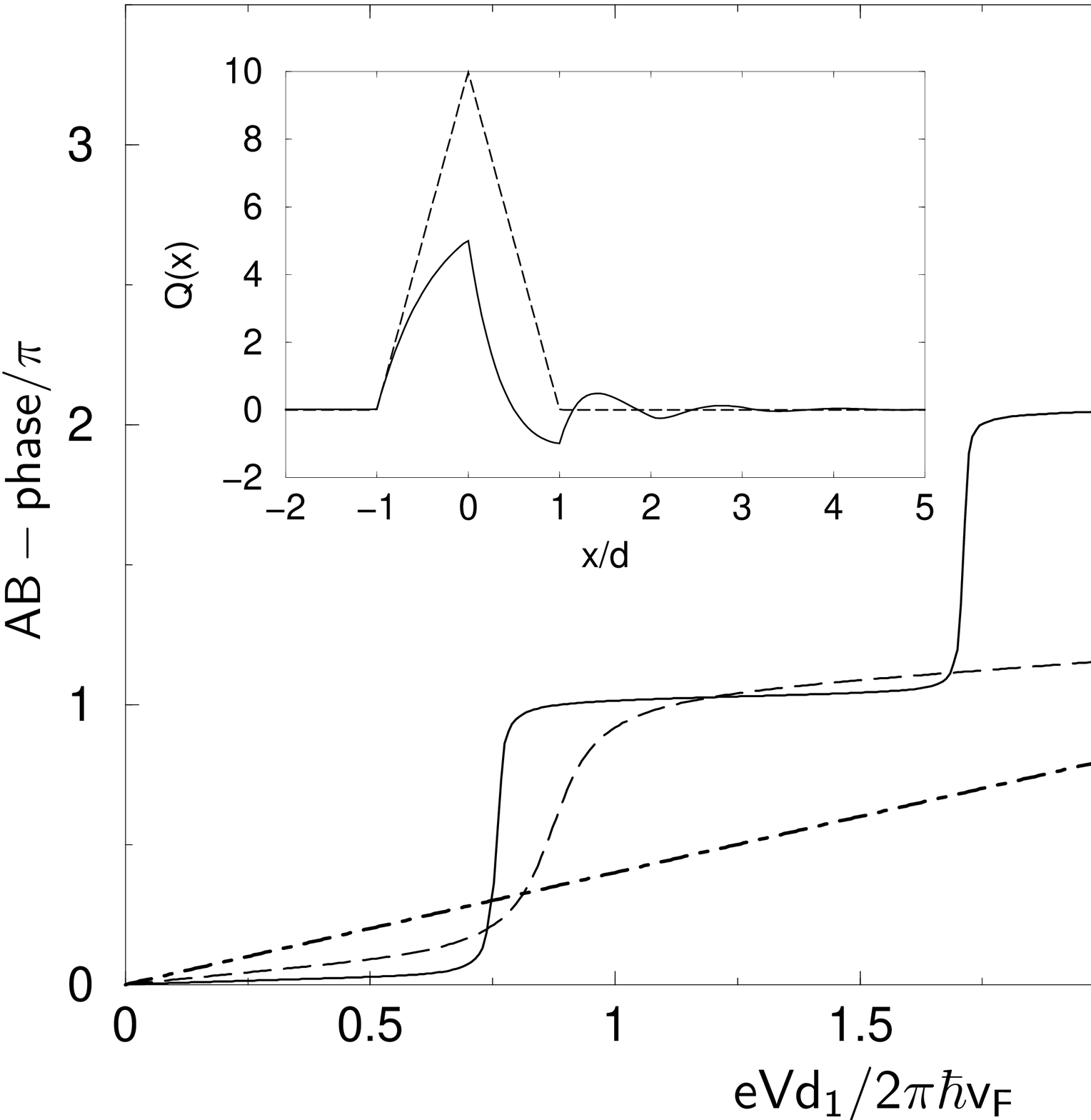,width=8cm} 
\caption{Dependence of interference fringe phase on bias voltage for
  MZI  with unequal arm lengths, $d_{2}/d_1 =1.2$, at interaction strengths:
$\tilde{\protect\gamma}=0$ (dot-dashed line),
$\tilde{\protect\gamma}=3$ (dashed line), and $\tilde{\protect\gamma}=10$ (full
line). Inset: the kernel $Q(x)$ of Eq.~(\ref{psi_dt}) at
$\tilde{\protect\gamma}=10$ (full line) and that of
Ref.~\cite{neder08} (dashed line).}
\label{fig3}
\end{figure}

In this Letter we present an exact calculation for a simplified
model of an interferometer. It reproduces the main signatures observed
experimentally \cite{heiblum2, roulleau07,bieri08}
and shows that the lobe pattern is a many-body effect, which would not appear in any approximation that treats single particles moving in a static mean-field potential. 
The model is illustrated in the inset to Fig.~\ref{fig2}. As in previous studies, two quantum Hall
edge channels, both with the same propagation direction, are coupled at two
quantum point contacts (QPCs). The simplifying feature of the model is that electrons
interact only when they are \textit{inside} the interferometer. This
allows us to combine a description of the contacts using fermion 
operators with a treatment of interactions using bosonization. Within the
MZI we take interactions only between two electrons on the same arm
and with fixed strength independent of distance, although it would be
feasible to relax these restrictions.
We consider an initial state in which Fermi seas in the
two channels are filled to different chemical potentials, to represent
the bias voltage, and evolve this state forward in time using the
Sch\"odinger equation. At long times the system reaches a stationary regime.
In this regime we calculate current and
differential conductance as a function of chemical potential difference and
enclosed AB flux. Our main results are presented in Figs.~\ref{fig2} and \ref
{fig3}, and discussed following an outline of their derivation; 
details will be presented elsewhere \cite{kovrizhin09prbmz}.

The solution we describe is significant more broadly as a rare example of a solved non-equilibrium scattering problem. One earlier instance is that of tunneling between fractional quantum Hall edge states \cite{fendley}, while another is the interacting resonant level model, treated recently by a form of Bethe Ansatz \cite{andrei}, and using boundary field theory \cite{boulat}. 
The remarkable structure observed experimentally \cite{heiblum2, roulleau07,bieri08} makes the MZI particularly interesting in this context. 

The Hamiltonian $\hat{H} = \hat{H}_{kin} + \hat{H}_{int} +
\hat{H}_{tun}$ for the model has three contributions, representing
respectively: kinetic energy, interactions, and tunneling at
contacts.  We formulate $\hat{H}$ initially for edges of length $L$ with periodic
boundary conditions, then take the limit $L\to\infty$. Then
\begin{equation}
\hat{H}_{kin}=-i\hbar v_{F}\sum_{\eta =1,2}\int_{-L/2}^{L/2}\hat{\psi}_{\eta
}^{+}(x)\partial _{x}\hat{\psi}_{\eta }(x)dx,  \label{H_kin}
\end{equation}%
where $v_{F}$ is the Fermi-velocity and $\eta =1,\,2$ is the channel
index. The Fermi field operators can be written as $\hat{\psi}%
_{\eta }(x)=L^{-1/2}\sum_{k}\hat{c}_{k\eta }e^{ikx}$, with
$k=2\pi n_{k}/L$ and $n_{k}$ integer, and
$\{\hat{c}_{k\eta },\hat{c}_{q\eta ^{\prime }}^{+}\}=\delta _{kq}\delta
_{\eta \eta ^{\prime }}$. Interactions are described by
\begin{equation}
\hat{H}_{int}=\frac{1}{2}\sum_{\eta =1,2}\int_{-L/2}^{L/2}U_{\eta
}(x,x^{\prime })\hat{\rho}_{\eta }(x)\hat{\rho}_{\eta }(x^{\prime
})dxdx^{\prime }\,,  \label{H_int}
\end{equation}%
where $\hat{\rho}_{\eta }\left( x\right) =\hat{\psi}_{\eta }^{+}(x)\hat{\psi}
_{\eta }(x)$ is the electron density operator. In our model $U_{\eta
}(x,x^{\prime })=0$ for $x,x^{\prime }\notin (0,d_{\eta })$.
Finally, the QPCs are represented by
\begin{equation}
\hat{H}_{tun}=v_{a}e^{i\alpha }\hat{\psi}_{1}^{+}(0)\hat{\psi}%
_{2}(0)+v_{b}e^{i\beta }\hat{\psi}_{1}^{+}(d_{1})\hat{\psi}_{2}(d_{2})+%
\mathrm{h.c.}  \label{H_tun}
\end{equation}%
The AB-phase appears here as $\varphi _{AB}\equiv \beta - \alpha $.

The total current $I$ from channel 1 to 2 has
contributions $I_a$ and $I_b$ arising from each QPC, which can be written in
terms of expectation values of operators acting at points
infinitesimally before the QPC. Each contribution can be separated
into a term that is not sensitive to coherence between the edges, and
another that is sensitive. We define $t_{a,b}=\sin \theta _{a,b}$ 
and $r_{a,b}=\cos \theta _{a,b}$ with $\theta
_{a,b}=v_{a,b}/\hbar v_{F}$, and 
denote expectation values by  $\langle \ldots \rangle$.
A straightforward
calculation yields for QPC $b$ the expressions $I_b = I_b^{(1)} +
I_b^{(2)}$, with
\begin{eqnarray} 
I_{b}^{(1)}&=&ev_{F}t_{b}^{2}\langle \hat{\rho}_{1}(d_1) -\hat{\rho}
_{2}(d_2)\rangle \nonumber\\ 
I_{b}^{(2)}&=&ev_{F}t_{b}r_{b}[ie^{i\beta }\langle \hat{G}_{12} \rangle +\mathrm{h.c.}]\;,  \nonumber
\end{eqnarray}
where $\hat{G}_{12}=\hat{\psi}_{1}^{+}(d_{1})\hat{\psi}
_{2}(d_{2}) $. Terms in $I_a$ are obtained from these for $I_b$ by
replacing $d_1$ and $d_2$ with $0$, and $v_b$ with $v_a$.
Since there is no coherence between channels before QPC $a$,
$I_a^{(2)}=0$ and the term responsible for AB oscillations in current
is $I_b^{(2)}$. The bias voltage is $V=(\mu_1-\mu_2)/e$ and the
differential conductance is $\mathcal{G}=e {\rm 
  d}I/{\rm d}\mu _{1}$
(with $\mu_2$ fixed). $\mathcal{G}$ oscillates with $\varphi _{AB}$, having
maximum and minimum values $\mathcal{G}_{\max }$ and
$\mathcal{G}_{\min }$,
and {\it AB fringe visibility} is defined as $(\mathcal{G}_{\max }-\mathcal{G}%
_{\min })/(\mathcal{G}_{\max }+\mathcal{G}_{\min })$.

The central task is therefore to calculate the correlator $\langle
\hat{G}_{12} \rangle$,
and our approach is as follows. (i)~We work in the interaction
representation, evolving operators with $\hat{H}_0 = \hat{H}_{kin} +
\hat{H}_{int}$ and treating $\hat{H}_{tun}$ as the `interaction'.
Then $\hat{\psi}_\eta\left( x,t\right) =e^{i\hat{H}_{0}t/\hbar}\hat{\psi}_\eta
\left( x\right) e^{-i\hat{H}_{0}t/\hbar}$ (note that we distinguish
operators in the Schr\"odinger and interaction representations by the
absence or presence of a time argument). The wavefunction of the
system, denoted at $t{=}0$ by $|Fs\rangle$, evolves with the S-matrix $\hat{S}\left( t\right)
=\mathrm{T}\exp \{-(i/\hbar)\int^{t}_0\hat{H} 
_{tun}\left( t^{\prime }\right) dt^{\prime }\}$, where
$\mathrm{T}$ indicates time ordering. (ii)~Time evolution of operators
is calculated using bosonization to diagonalise
$\hat{H}_0$. (iii)~Results are written in terms of operators in the Schr\"odinger
picture, with boson operators re-expressed using fermion ones.
This yields an expression for $\hat{G}_{12}$ suitable for straightforward numerical
evaluation. We next outline these three steps.

Step (i): Evaluation of $\hat{S}(t)$ hinges on our restriction of interactions
to the interior of the MZI. Specifically, separating $\hat{H}_{tun}$
into parts  $\hat{H}_{tun}^a$ and  $\hat{H}_{tun}^b$ due to each QPC,
we find from step (ii) that
$[\hat{H}_{tun}^a(t_1),\hat{H}_{tun}^b(t_2)]=0$ 
and $[\hat{G}_{12}(t_1),\hat{H}^b(t_2)]=0$, provided $t_1 \geq
t_2$. The first commutator allows us to factorise the S-matrix as $\hat{S}(t) = \hat{S}^b(t)
\hat{S}^a(t)$, where $\hat{S}^a(t)$ is calculated using $\hat{H}_{tun}^a$ and
$\hat{S}^b(t)$ using $\hat{H}_{tun}^b$. The second ensures that
$[\hat{S}^b(t)]^+\hat{G}_{12}(t)\hat{S}^b(t) = \hat{G}_{12}(t)$, so that an explicit form for $\hat{S}^b(t)$ is not
required in the calculation. Since QCP $a$ acts before interactions,
$\hat{S}^a(t)$ is easy to evaluate: we have
$[\hat{H}_{tun}^a(t_1),\hat{H}_{tun}^a(t_2)]=0$ for any $t_1,t_2 \geq
0$ and so may omit time ordering. In particular, we will need to compute the action of
$\hat{S}^a(t)$ on fermionic operators. It is a rotation in the space
of channels and can be written
$\tilde{\hat{\psi}}_{\eta }(x)=[\hat{S}^a(t)]^{+}\hat{\psi}_{\eta^{\prime}
}(x)\hat{S}^a(t)$. For $0<x< v_F t$ we find
\begin{eqnarray}
\tilde{\hat{\psi}}_{\alpha }(x)&=&\sum_{\beta }{\cal S}_{\alpha \beta }^{a}\hat{\psi}%
_{\beta }(x),\nonumber\\ {\cal S}^{a}&=&\left( 
\begin{array}{cc}
r_{a} & -it_{a}e^{i\alpha } \\ 
-it_{a}e^{-i\alpha } & r_{a}%
\end{array}%
\right)\;.  \label{Sa}
\end{eqnarray}

Step (ii): We compute time evolution under $\hat{H}_0$ using
bosonization \cite{vonDelft}. Fermion operators are written in the form 
\begin{equation}
\hat{\psi}_{\eta }(x)=(2\pi a)^{-1/2}\hat{F}_{\eta }e^{i\frac{2\pi }{L}\hat{N
}_{\eta }x}e^{-i\hat{\phi}_{\eta }(x)},  \label{boson_id}
\end{equation}
where $\hat{F}_{\eta }$ are Klein factors with commutation relations $
\{\hat{F}_{\eta },\hat{F}_{\eta ^{\prime }}^{+}\}=2\delta _{\eta \eta
^{\prime }}$ and bosonic fields are defined as 
\begin{equation}
\hat{\phi}_{\eta }\left( x\right) =-\sum_{q>0}\left( 2\pi /qL\right)
^{1/2}(e^{iqx}\hat{b}_{q\eta }+\mathrm{h.c.})e^{-qa/2},  \label{phi_x}
\end{equation}%
with $a$ an infinitesimal regulator. Plasmon creation operators obey bosonic
commutation relations $[\hat{b}_{q\eta },\hat{b}_{k\eta ^{\prime
}}^{+}]=\delta _{qk}\delta _{\eta \eta ^{\prime }}$ and are expressed
for $q>0$ in terms
of fermions as 
\begin{equation}
\hat{b}_{q\eta }^{+}=i\left( 2\pi /qL\right) ^{1/2}\sum_{k=-\infty}^{\infty}\hat{c}_{k+q\eta
}^{+}\hat{c}_{k\eta } \;. \label{b_op}
\end{equation}%
Since $\hat{H}_0$ does not couple channels, we restrict attention to a
single channel and omit channel labels until we reach step~(iii).
The kinetic energy $\hat{H}_{kin}$ for a single edge has the
bosonized form
\begin{equation}
\hat{H}_{kin} =\frac{\hbar v_{F}}{2}\int_{-L/2}^{L/2}\frac{dx}{2\pi }%
(\partial _{x}\hat{\phi}\left( x\right) )^{2}+\frac{2\pi }{L}\frac{\hbar
v_{F}}{2}\hat{N}(\hat{N}+1)
\end{equation}
where $\hat{N}\equiv \sum_{k}\hat{c}_{k}^{+}\hat{c}_{k}$ is the particle
number operator. Similarly, $\hat{H}_{int}$ is quadratic when written
using the bosonic representation of the density operators, $\hat{\rho}\left( x\right) =-\frac{1
}{2\pi }\partial _{x}\hat{\phi}\left( x\right) +\hat{N}/L$. 
The time dependence of $\hat{\phi}\left( x,t\right)$ can be found by solving the equation of motion. 
Since our choice of non-uniform interactions leads to a
coupling between the plasmon and number operators, we make the separation
$\hat{\phi}\left( x,t\right)= \hat{\phi}^{(0)}\left( x,t\right)+\hat{\phi}^{(1)}\left( x,t\right)$, where
$\hat{\phi}^{(0)}\left( x,t\right) \propto \hat{N}/L$ and $\hat{\phi}^{(1)}\left( x,t\right)$
is independent of $\hat{N}$, satisfying
\begin{equation}
2 \pi \hbar (\partial_t + v_F \partial_x) \hat{\phi}^{(1)}\left( x,t\right) =  -\int U(x,y) \partial_y \hat{\phi}^{(1)}\left( y,t\right) {\rm d}y\,.
\end{equation}
The solution can be written in the form
\begin{equation}
\hat{\phi}^{(1)}\left( x,t\right) = \int_{-L/2}^{L/2}  K(x,y;t)[ \hat{\phi}\left( y\right) -\hat{\phi}^{(0)}(y)]{\rm d}y\,,\nonumber
\end{equation}
where the Green function $K(x,y;t)$ can be constructed in the usual way from the eigenfunctions of the time-independent equation,
\begin{equation}
2 \pi \hbar v_F(\partial_x - ip) f_p(x) =  -\int U(x,y) \partial_y f_p(y) {\rm d}y\,.\nonumber
\end{equation}

We now specialise to interactions that are constant within the
interferometer: $U(x,x^{\prime })=g$ for $x,x^{\prime
}\in (0,d)$ and $U(x,x^{\prime })=0$ otherwise. This form of the
potential is the one
treated approximately in \cite{neder08}. It is
characterised by the dimensionless coupling constant $\gamma =gd/2\pi
\hbar v_{F}$. We
find in the limit $L \to \infty$
\begin{equation*}
f_{p}\left( x\right) =
\left\{ 
\begin{array}{ll}
e^{ipx} & x \leq0 \\ 
r_{p}+s_{p}e^{ipx} & 0<x<d \\ 
e^{ipx-i\delta _{p}} & x \geq d%
\end{array}%
\right. \;.
\end{equation*}%
The coefficients $s_{p}=(1+t_{p})^{-1}$ and $r_{p}=t_{p}s_{p}$, with $t_{p}=(i\gamma/
p d)(1 - e^{ipd})$, are obtained from
matching 
$f_{p}(x)$ at $x=0,d.$ The phase shifts of plasmons $\delta _{p}$ due to the
interactions are given by 
$e^{-i\delta _{p}}=(1+t_{p}^{\ast })/(1+t_{p})$. 
Similarly, we find $\hat{\phi
}^{(0)}\left( x\right) =2\pi \bar{\gamma}\hat{N}x/L$ for $x\in (0,d),$
where $\bar{\gamma}=\gamma (1+\gamma )^{-1}$. 

In this way we find an expression for $K(x,y;t)$. Setting $x=d$, it simplifies at long times to
\begin{equation}
K(d,y;t) = \frac{1}{2\pi} \int_{-\infty}^{\infty} {\rm d} p\, e^{i(p[d-y-v_Ft] - \delta_p)}\,.
\end{equation}
Using this and Eqns.~(\ref{phi_x}) and (\ref{b_op}), we write $\hat{\phi}^{(1)}\left( x,t\right)$ as a bilinear in the fermion operators $\hat{c}^+_k$ and $\hat{c}_k$.

Step (iii): We employ this result to construct an expression for
$\hat{G}_{12}(t)$ in terms of fermion operators 
in the Schr\"odinger representation. To this
end, we start from Eq.~(\ref{boson_id}) in the interaction representation at
time $t$ and substitute for $\hat{\phi}_\eta(d_\eta,t)$ as described.
We also eliminate the combination ${\cal F}
\equiv (2\pi a)^{-1/2}\hat{F}_{\eta }e^{i\frac{2\pi }{L}\hat{N
}_{\eta }d_\eta}$ by inverting the bosonization identity,
Eq.~(\ref{boson_id}), writing
\begin{equation}
{\cal F}(t) 
= 
e^{i\hat{H}_{kin}t/\hbar}
{\cal F} e^{-i\hat{H}_{kin}t/\hbar} = 
\hat{\psi}_{\eta }(z)e^{i\hat{\phi}_{\eta }(z)}\nonumber
\end{equation}
for $z_\eta=d_\eta - v_Ft$. Finally, we substitute for $\hat{b}_{q \eta}$ and
$\hat{b}_{q \eta}^+$ in $\hat{\phi}_\eta(x)$
using Eq.~(\ref{b_op}). The result
(omitting an unimportant, constant phase) is the operator identity
\begin{equation}
\hat{\psi}_\eta(d_\eta,t)=e^{-i\hat{Q}_\eta}\hat{\psi}_\eta(z_\eta)\;.  \label{psi_dt}
\end{equation}%
Here $\hat{Q}_\eta = \int_{-\infty }^{\infty }Q_\eta(x-z_\eta)\hat{\rho}_\eta(x)
dx$, where the kernel $Q_\eta(x)=L^{-1}\sum_{q=-\infty }^{\infty }\tilde{Q}_\eta(q)e^{iqx}$ 
has for our choice of interaction the Fourier transform
\begin{equation}
\tilde{Q}_\eta(q)=2\pi \gamma d_\eta j_{0}^{2}(qd_\eta/2)(1+ \gamma e^{-iqd_\eta/2}j_{0}(qd_\eta/2))^{-1},
\label{kernel}
\end{equation}%
in which $j_{0}(x)=x^{-1}\sin x$.

In this way we arrive at the expression
\begin{equation}
\langle G_{12}(t) \rangle = e^{i\bar{\Phi}}\langle Fs|[\hat{S}^a(t)]^{+}\hat{\psi}
_{1}^{+}(z_1)e^{i\hat{R}}\hat{\psi}_{2}(z_{2})\hat{S}^a(t)\left\vert
Fs\right\rangle\;.\nonumber
\end{equation}
Here 
$\bar{\Phi}$ is an initial phase that is
independent of voltage, and $\hat{R}=\hat{Q}_1 - \hat{Q}_2$. 
The action of $\hat{S}^a(t)^{+}$
and $\hat{S}^a(t)$ on the operators they enclose is given by
Eq.~(\ref{Sa}), 
and evaluation of $\langle\hat{G}_{12}(t)\rangle$ reduces to the calculation
of correlators of the form $C_{\mu\eta}=\langle Fs|\hat{c}_{\mu}%
^{+}\exp({i\sum_{\alpha \beta}\mathrm{M}_{\alpha \beta}\hat{c}_{\alpha}^{+}\hat{c}_{\beta}})c_{\eta}|Fs\rangle
$, where the indices specify both channel and momentum, and the matrix
$\mathrm{M}$ is obtained from $[\hat{S}^{a}(t)]^{+}\hat{R}\hat{S}^{a}(t)$. One
can show that $C_{\mu\eta}=\mathrm{{D}}_{\eta\mu}^{-1}\det
\mathrm{{D}}$ with $\mathrm{{D}}$ constructed from the matrix
elements of $\exp(i\mathrm{M})$ between the single-particle states
that are occupied in the Slater determinant $|Fs\rangle$. We
calculate $C_{\mu\eta}$ numerically, achieving 
convergence of the results when keeping up to $10^{3}$ basis states and $400$
particles in each channel.

The physical interpretation of the solution we have presented is as
follows. Each electron passing QPC $b$ at time $t$ has an accumulated
phase from its interactions with other electrons. The phase
is a collective effect and is represented by the operator
$\hat{Q}_\eta$ in Eq.~(\ref{psi_dt}).
Contributions from interactions with particles at a distance $x$ 
from the one at QPC $b$ have a weight determined by the kernel
$Q_\eta(x)$, illustrated in the inset to Fig.~\ref{fig3}. 
This weight is largest near $x=0$, showing that interactions with
nearby electrons are most important. Moreover, since $Q_\eta(x) = 0$ for
$x<-d_\eta$, a given electron is uninfluenced by the ones behind,
that enter the interferometer after it exits. The precise form of
the kernel reflects the full many-body physics of the problem:
a similar kernel appears in Eq.~(11) of Ref.~\cite{neder08},
but with a simpler form because of the approximations employed
there.

A consequence of the phase $\hat{Q}_\eta$ is that many-particle
interference influences the MZI conductance. As an illustration, consider
the quantum amplitudes for two particles to pass through the
interferometer on all possible paths connecting given initial and final
states. Paths for which both particles propagate on the same arm of the
interferometer have an interaction contribution to their phase that
varies with their separation and
is absent if the two particles propagate on different arms. Destructive
interference between paths with different interaction phases generates the observed
lobe structure.

We now turn to our results. The parameters in the model are: the dimensionless interaction
strength $\gamma$, the transmission probabilities
$t_a^2$ and $t_b^2$, the ratio $d_2/d_1$ of arm lengths, and the
dimensionless bias voltage $eV\sqrt{d_1 d_2}/2\pi \hbar v_F$. 
We consider $1\leq 2\pi \gamma \leq 10$, $1\leq d_2/d_1 \leq 1.2$ and
first discuss behaviour with $t_a^2 = t_b^2 = 1/2$.
The dependence of visibility of AB fringes on bias voltage
and interaction strength is presented in Fig.~\ref{fig2}, taking
equal arm lengths and transmission probabilities of
$1/2$ at both QPCs. The key features of all three curves in this
figure match those of the experiment (see Figs.~2 and 3 of
\cite{heiblum2}): with increasing bias there is a sequence of lobes in
the visibility, which have decreasing amplitude and are separated by
zeros. The phase of AB fringes is also influenced by
interactions. Results are displayed in Fig~\ref{fig3}. For an
MZI with different arm lengths (as in this figure), the fringe phase
without interactions varies linearly with bias, because the Fermi
wavevector $k_F$ is linear in bias and the phase difference between
particles traversing the two arms is $k_F(d_2 - d_1)$. With increasing
interaction strength the phase dependence on bias develops into a
series of smooth steps, each of height $\pi$. The risers of these
steps coincide with minima of the visibility. Strikingly, with strong
interactions phase steps at minima of the visibility persist for
$d_1=d_2$, even though in this case phase would be independent of bias
without interactions.  The stepwise phase
variation we find at large interaction strength also matches
observations (see Fig.~2 of \cite{heiblum2}).

Behaviour is insensitive to the transmission probability $t_b^2$ at
QPC $b$, apart from the overall scale for visibility. Departures from
$t_a^2=1/2$, however, eliminate the exact zeros in visibility, leaving
only sharp minima. 
A difference in arm lengths has a similar though much weaker effect.

The width in bias voltage of the central visibility lobe defines an
energy scale. In our model this scale is of order $g$ at large
$\gamma$. Taking $v_F = 2.5\times 10^{4} {\rm ms^{-1}}$, $d=10\mu {\rm
  m}$ and the permittivity $\epsilon =12.5$ of GaAs, we estimate from
the capacitance of an edge channel $g\sim 10{\rm \mu eV}$. This is
similar to the experimentally observed value of about $14\ \mathrm{\mu
  eV}$ \cite{heiblum2}.

Our calculations rely on a simplified form for 
interactions, but we believe our choice is quite reasonable.
Our central approximation is to neglect interactions between
an electron inside the MZI and one outside. In practice, such
interactions will anyway be screened by the metal gates that define
the QPCs. We also neglect interactions between a pair of electrons
that are both outside the MZI. This is unimportant: before
electrons reach the MZI, such interactions do not cause scattering
because of Pauli blocking, while after electrons pass through the MZI,
these interactions cannot affect the current. 
Within the MZI we represent interactions by a
capacitative charging energy. Such a
choice is standard in the theory of quantum dots and has been
applied previously to interferometers \cite{buttiker01,neder08}.

In summary, we have calculated the visibility of
Aharonov-Bohm fringes in the differential conductance of an electronic
MZI out of  equilibrium, taking exact account of
interactions between electrons.
From our calculations
we obtain a lobe pattern in the dependence of visibility
on bias, and jumps in the phase of fringes at zeros of the visibility, as observed
experimentally \cite{heiblum2, roulleau07,bieri08}. 

We thank F. H. L. Essler for fruitful discussions and acknowledge support from
EPSRC grants EP/D066379/1 and EP/D050952/1.


\begin{thebibliography}{99}
\bibitem{heiblum2} I. Neder {\it et al.},
Phys. Rev. Lett. \textbf{96}, 016804, (2006). 

\bibitem{roulleau07} P. Roulleau {\it et al.},
Phys. Rev. B \textbf{76},
161309(R) (2007). 

\bibitem{bieri08} E. Bieri {\it et al.},
Phys. Rev. B {\bf 79}, 245324 (2009).

\bibitem{glazman} A. Imambekov and L. I Glazman, Science {\bf
        323}, 228 (2009)

\bibitem{fractional}
C. de C. Chamon {\it et al}, 
Phys. Rev. B \textbf{55}, 2331 (1997); 
F. E. Camino, W. Zhou, and V. J. Goldman, {\it ibid.} {\bf 72}, 075342
(2005); K. T. Law, D. E. Feldman, and Y. Gefen, {\it ibid.}
\textbf{74}, 045319 (2006); D. E. Feldman and A. Kitaev, Phys. Rev. Lett. \textbf{\
97}, 186803 (2006); V. V. Ponomarenko and D. V. Averin, {\it ibid.}
\textbf{99}, 066803 (2007). 

\bibitem{marquardt04} F. Marquardt and C. Bruder, Phys. Rev. Lett. \textbf{92%
}, 056805 (2004).

\bibitem{buttiker01} G. Seelig and M. B\"uttiker, Phys. Rev. B \textbf{\ 64}%
, 245313 (2001).



\bibitem{chalker07} J. T. Chalker, Y. Gefen, and M. Y. Veillette, Phys.
Rev. B \textbf{76} 085320 (2007). 

\bibitem{sukhorukov07} E. V. Sukhorukov and V. V. Cheianov, Phys. Rev. Lett. 
\textbf{99}, 156801 (2007).

\bibitem{sukhorukov08} I. P. Levkivskyi, E. V. Sukhorukov, Phys. Rev. B 
\textbf{78}, 045322 (2008).

\bibitem{neder08} I. Neder and E. Ginossar, Phys. Rev. Lett. \textbf{100},
196806 (2008).

\bibitem{sim08} Seok-Chan Youn, Hyun-Woo Lee, and H.-S. Sim, Phys. Rev.
Lett. \textbf{100} 196807 (2008). 

\bibitem{marquardt08} 
B. Abel and F. Marquardt, Phys. Rev. B {\bf 78}, 201302 (2008).


\bibitem{kovrizhin09prbmz} D.L. Kovrizhin and J. T. Chalker, in preparation.

\bibitem{fendley} P. Fendley, A. Ludwig, and H. Saleur, Phys. Rev. Lett. {\bf  75}, 2196 (1995). 

\bibitem{andrei} P. Mehta and N. Andrei, Phys. Rev. Lett. {\bf 96}, 216802 (2006).

\bibitem{boulat} E. Boulat, H. Saleur, and P. Smitteckert, Phys. Rev. Lett. {\bf 101}, 140601 (2008). 

\bibitem{vonDelft} See: J. von Delft and H. Schoeller, Annalen Phys. \textbf{7},
225 (1998); T. Giamarchi, {\it Quantum Physics in One Dimension}
(Oxford Univ. Press, Oxford, 2004).  



\end{thebibliography}
\end{document}